 \documentclass{ws-ijmpa}
\usepackage{color}
\usepackage[super,compress]{cite}
\usepackage{graphicx}
\begin{document}
\markboth{D. Blaschke et al.}{Thermal production of sexaquarks in heavy-ion collisions}

%
\catchline{}{}{}{}{}
%

\title{Thermal production of sexaquarks in heavy-ion collisions}

\author{David Blaschke}
\address{Institute of Theoretical Physics, University of Wroclaw, Max Born Pl. 1. 50-204 Wroclaw, Poland}
\address{Bogoliubov Laboratory of Theoretical Physics, JINR Dubna, Joliot-Curie Str. 6, 141980 Dubna, Russia}
\address{National Research Nuclear University (MEPhI), Kashirskoe Shosse 31, 115409 Moscow, Russia}
\author{Larissa Bravina}
\address{University of Oslo, POB 1048 Blindern, N-0316 Oslo, Norway}
\author{Kyrill Bugaev}
\address{Bogolyubov Institute for Theoretical Physics, Metrologichna str. 14B, Kyiv 03680, Ukraine}
\address{Department of Physics, Taras Shevchenko National University of Kyiv, 03022 Kyiv, Ukraine}
\author{Glennys R. Farrar}
\address{Department of Physics, New York University, New York, NY, USA}
\author{Boris Grinyuk}
\address{Bogolyubov Institute for Theoretical Physics, Metrologichna str. 14B, Kyiv 03680, Ukraine}
\author{Oleksii Ivanytskyi}
\address{Institute of Theoretical Physics, University of Wroclaw, Max Born Pl. 1. 50-204 Wroclaw, Poland}
\address{Bogolyubov Institute for Theoretical Physics, Metrologichna str. 14B, Kyiv 03680, Ukraine}
\author{Sonia Kabana\footnote{Corresponding author Sonja.Kabana@cern.ch}}
\address{Instituto de Alta Investigaci\'on, Universidad de Tarapac\'a , Casilla 7D, Arica, Chile}
\author{Sergey V. Kuleshov}
\address{Universidad Andr\'es Bello, Avenida Republica 237, Santiago,  Chile and Millennium Institute for SubAtomic Physics at the High-energy FrontIer (SAPHIR) of ANID, Chile.}
\author{Irina K. Potashnikova}
\address{Departamento de F\'{\i}sica, Universidad T\'ecnica Federico Santa Mar\'ia, Avenida Espa\~na 1680, Valpara\'iso, Chile}
\author{Violetta Sagun}
\address{CFisUC, Department of Physics, University of Coimbra, Coimbra, 3004-516 Portugal}
\address{Bogolyubov Institute for Theoretical Physics, Metrologichna str. 14B, Kyiv 03680, Ukraine}
\author{Arkadiy Taranenko}
\address{National Research Nuclear University (MEPhI), Kashirskoe Shosse 31, 115409 Moscow, Russia}
\author{Oleksandr V. Vitiuk}
\address{Department of Physics, Taras Shevchenko National University of Kyiv, 03022 Kyiv, Ukraine}
\author{Evgeny Zabrodin}
\address{University of Oslo, POB 1048 Blindern, N-0316 Oslo, Norway}
\address{Skobeltzyn Institute of Nuclear Physics (SINP), Moscow State University, RU-119899 Moscow, Russia}
\author{Xiaoming Zhang}
\address{Central China Normal University, Wuhan, Hubei 430079}
\author{Daicui Zhou}
\address{Central China Normal University, Wuhan, Hubei 430079}
\address{Quark and Lepton key Lab. (Ministry of Education), Institute of Particle Physics Central China Normal University, Wuhan, Hubei 430079}

\maketitle

\begin{abstract}
We present new results on the thermal production yield
of a hypothetical state made of six quarks $uuddss$ assuming 
its production in heavy-ion collisions at the CERN Large Hadron  Collider (LHC).
A state with this quark content
and mass low enough to be stable against decay in 
timescales of the order of the age of the Universe, has been
hypothesized by one of us (G.F.) and has been discussed as
a possible  dark matter candidate. 
In this work we address for the first time 
the thermal production rate that can be expected for this state in heavy-ion collisions at colliders. 
For this estimate we use a thermal model which has been shown 
to describe accurately the production of hadrons and nuclei 
in heavy-ion collisions at LHC energy. 
This estimate is of great relevance for sexaquark searches at colliders as well as for its consideration as a dark matter candidate and for the composition of neutron stars.

\end{abstract}

\ccode{PACS numbers:}

\section{Introduction}

The papers on quarks by Gell-Mann and Zweig opened the possibility that hadrons with more than the minimal quark content ($q\bar{q}$ or $qqq$) could exist 
\cite{Zweig:1981pd,GellMann:1964nj}.
In recent years experiments have discovered some multiquark states, demonstrating  that such states do exist in nature 
and they are currently included in the Review of Particle Properties\cite{PDG}.
Among them we mention as an example the first observation of pentaquark states by the LHCb Collaboration
that was not
opposed by other experiments via a non-observation with Upper limits conflicting the measured
cross sections 
\cite{LHCb_pentaquarks, LHCb_pentaquarks2}.
In the analysis of $\Lambda_b^0 \rightarrow J/\psi \,  \, p \, K- $  
decays they observed narrow  peaks in the $J/\psi \,  \, p$ mass
spectrum  and further analysis 
led to the discovery of 
the $P_c(4380),  P_c(4440), P_c(4457)$ and $P_c(4312)$ \cite{LHCb_4312}  pentaquark states with 
valence quark content $uuds \overline{c}$, which are listed in \cite{PDG}.

Furthermore, the LHCb Collaboration reported in 2021 
a structure seen in the $J/\psi \,  \Lambda$ invariant mass distribution
at a mass of 4458.8$\pm$ 2.9 +4.7 - 1.1 MeV (with statistical and syst. errors, respectively) 
 obtained from an amplitude analysis of  $ \Xi_b^- \rightarrow J/\psi \,  \Lambda K^-$ decays
\cite{LHCb_JpsiLambda_4458}.
To mention a recent result on tetraquarks from another experiment,
in 2021 the BESIII collaboration reported
 the discovery of an exotic tetra-quark structure,the $Z_{cs}$(3985), produced in 
$ e^+ e^- \rightarrow K^+ (D_s- D^*_0+D_s^*-D_0)$  at center-of-mass energy of 4.68 GeV
and has a minimal $c \overline{c} s \overline{u}$ four-quark substructure
\cite{BESIII_3985}.
Apart from the above given few examples from experimental findings on multiquark states,
for a review on discovered multiquark states or multiquark state candidates
we refer to the reviews for exotic hadrons in the PDG \cite{PDG} and to the
 review  \cite{review_Marek}.

Among the  multiquark  states the neutral flavor singlet 
multiquark state $uuddss$ has been long discussed as an interesting multiquark candidate, because it is expected to be tightly bound, as first discussed by R. Jaffe \cite{Jaffe:1976yi}.
Using  a bag model, R. Jaffe estimated that the $uuddss$ state is expected to have a mass around 2150 MeV, making it strong interaction stable but allowing 1st-order weak
decay into $\Lambda + p + \pi^-$.
R. Jaffe named the $uuddss$ candidate with
such characteristics, the $H0$ dibaryon.
A number of experiments have searched for such a state and did not find it.
But they have set upper limits to its existence,
see, e.g.,  Refs.\cite{Kim:2013vym,Badier:1986xz,Bernstein:1988ui,Belz:1995nq,AlaviHarati:1999ds,Gustafson:1976hd}.

It has been recently proposed by one of us (G.F.), that a color-flavor-spin-singlet multiquark state made of $uuddss$  quarks may exist, with a hard-core radius of $\sim$ 0.1-0.4 fm
and  mass  lower than  2 GeV, in which case it would be stable 
or have a lifetime greater than the age of the Universe
\cite{1708.08951, 1711.10971, 1805.03723}.
In addition, such a $uuddss$ state would couple weakly to hadrons because of its flavor singlet nature that excludes coupling to pions \cite{1708.08951, 1711.10971, 1805.03723}.

In that case, because of its weak coupling matter
and its stability, the $uuddss$ state has escaped observation 
in experiments till now, since experiments in the past searched
for the $H0$ dibaryon with the characteristics as discussed by 
R. Jaffe.
Therefore, the upper limits set by these experiments do not
apply to a $uuddss$ state with the above discussed characteristics \cite{1708.08951, 1711.10971, 1805.03723}.

For the above reasons the $uuddss$ state with such characteristics was proposed by one of us (G.F.) as a possible dark matter candidate
\cite{1708.08951, 1711.10971, 1805.03723}.
To distinguish the low mass stable and compact $uuddss$ state
from the weakly decaying $uuddss$ state of R. Jaffe,
G. Farrar proposed the name sexaquark ($S$).

As discussed by G. Farrar in \cite{1805.03723}, assuming that  dark matter is made of equal numbers of $u,d$ and $s$ quarks 
(in any form, namely it could be a small object as a sexaquark but also a big one like a quark star)
leads to an estimated ratio of dark matter to ordinary matter which agrees within a factor of $2$ with the measured
ratio of dark matter to ordinary matter in the Universe, for the whole range of masses shown in Fig.~2 of \cite{1805.03723}\,;
namely 1850-1890 MeV and for quantum chromodynamics (QCD)  phase transition  temperature
of 140-170 MeV.  
This is a  very interesting similarity of predicted and observed values of the dark matter to baryon abundances,   
suggesting that maybe objects made out of $u,d,s$ quarks can be at the origin of dark matter in the Universe
since there is a priori no reason  why the numbers shown in 
Fig.~2 of \cite{1805.03723} agree with each other, while they could be different by many orders of magnitude 
\footnote{As shown in subsequent work, within the uncertainties of the calculation and nature of the evolution from the quark-gluon plasma to the hadronic phase as chiral symmetry breaking and confinement turn on, the range of sexaquark masses which can be compatible with the observed dark matter to baryon ratio is larger, possibly from 1700 MeV to 2 GeV.  Note a minor plotting error in \cite{1805.03723} which is corrected in \cite{fwx20_v2}. }.
We note that the recent result for $T_{c}$ of the QCD phase transition, obtained from recent simulations of lattice QCD is $156.5 \pm 1.5$ MeV  \cite{lattice}. However, the small statistical error of this result can not be mismatched with the width of the crossover transition region that can be estimated from the full width at half maximum of the chiral susceptibility and which is at least one order of magnitude larger.

Questions have been raised in the literature as of the possibility that the sexaquark can be a candidate for dark matter (see, e.g., \cite{Kolb_Turner}).
However, this possibility is not excluded since the authors of \cite{Kolb_Turner} neglected to take into account the robustness of sexaquarks in the hot hadronic phase against breakup to baryons, which follows if the sexaquark is very compact and/or there is a high potential barrier for reconfiguring the quarks.
 
Experiments aim to explore all possibilities to discover dark matter and started searching for a possible stable sexaquark.  In particular the BABAR collaboration was the first experimental collaboration that performed a search for sexaquarks and
published an upper limit on its production in Upsilon decays.
\cite{BABAR}.
In this search they explored the detection method proposed in 
\cite{1708.08951}, namely to search for the $S$ state via the Upsilon decay in Upsilon factories, as missing mass due to the 
$S$ or $\bar{S}$ production in association with di-anti-Lambda or di-Lambdas: 
$ \Upsilon \rightarrow 
  S \overline{\Lambda} \overline{\Lambda} 
$
or
$
\overline{S} \Lambda \Lambda
$
+ pions and/or photons.
The BABAR collaboration observed no signal in the exclusive  
$S \overline{\Lambda} \overline{\Lambda} $ + charge conjugate channel and published the 90\% confidence level limits on the branching fraction 
$\Upsilon(2S,3S) \rightarrow  S \overline{\Lambda} \overline{\Lambda} $  to be (1.2-1.4) $10^{-7}$
for a mass of the $S$ below 2.05 GeV.

Comparing this result to expectations, the predicted
\cite{1708.08951} {\it inclusive} branching fraction of $S$ in Upsilon decays is of the order $\sim 10^{-7}$.   
The exclusive to inclusive ratio for Upsilon decay is typically $\lesssim 10^{-4}$, so no exclusive signal would have been expected at the level of sensitivity of BABAR which is $10^{-7}$
in the exclusive channel as discussed in \cite{1708.08951}.

In the present paper we estimate for the first time the  production yield of the sexaquark in heavy-ion collisions
at the LHC assuming a thermal production of the sexaquark. 

We use a model for thermal production of hadrons and nuclei in heavy-ion collisions which has successfully described data at the LHC and other collision energies \cite{KAB_Ref4,KAB_Ref5,KAB_Ref6new}.
In section \ref{section_model} we describe the model and in section \ref{section_results} we present the results and their discussion ending in section \ref{section_conclusions} with conclusions and outlook.

\section{The hadron resonance gas model (HRGM) with multicomponent hard-core repulsion}
\label{section_model}

The hadron resonance model with induced surface tension which allows us to treat the hard-core repulsion of hadrons and light (anti-)nuclei has been developed by some of us 
and is discussed in detail in references 
\cite{KAB_Ref4,KAB_Ref5,KAB_Ref6new,KAB_Ref14,KAB_Ref15,KAB_Ref16}. 
The model was designed for an accurate description of multicomponent hadron mixtures. 
The effects of hard-core repulsion are known to be important for such systems. 
Within the present approach their treatment beyond the Van der Waals approximation is provided by attributing a part of the interparticle repulsion to the induced surface tension coefficient $\Sigma$. 
It is remarkable, that regardless the number of particle species in the grand canonical ensemble this quantity is defined along with the system pressure $p$ as a solution of the system of two coupled equations (we use the natural units, i.e. speed of light $c=1$, the Planck constant $\hbar=1$ and the Boltzmann constant $k_B=1$ are set to a unity).

\begin{eqnarray}
\label{EqI}
p&=& T \sum_{k=1}^N \phi_k \exp \left[ 
\frac{ - v_k p - s_k \Sigma}{T} \right]\,,
\\ 
\label{EqII}
\Sigma&=&
T \sum_{k=1}^N R_k \phi_k \exp 
\left[ \frac{- v_k p - s_k \alpha \Sigma}{T} \right] \,.
\end{eqnarray}

The summation in this expressions is performed over all particle species labelled by $k$. 
Each of them is assigned an own proper volume $v_k = \frac{4}{3}\pi R_k^3$ and surface area $s_k = 4\pi R_k^2 $ expressed through the corresponding hard-core radius $R_k$. 
The parameter $\alpha$ plays a crucial role within the present approach. It accounts for the higher order virial coefficient of the equation of state (\ref{EqI}) - (\ref{EqII}). 
In Refs.~\cite{KAB_Ref4,KAB_Ref5} it was found that in the limit of Boltzmann statistics, $\alpha = 1.25$ reproduces the third and the fourth virial coefficient of the gas of hard spheres, while the widely used Van der Waals prescription is consistent only with the second one. It is important to note, that the description of hadronic mixtures with the Boltzmann statistics is well justified at the LHC energies since the corresponding particle occupation numbers are high. 
Accounting for this fact, we can write the  thermal density of the $k$-th hadron species as
\begin{eqnarray}
\label{EqIII}
\phi_k&= &\frac{g_k}{N_k (M_k^{Th})}   
\int\limits_{M_k^{Th}}^\infty  \, 
\frac{dm~\Gamma_{k}}{(m-m_{k})^{2}+\Gamma^{2}_{k}/4}
\times
\int \frac{d^3 p}{ (2 \pi)^3 }   \exp \left[\frac{\mu_k- \sqrt{p^2 + m^2} }{T} \right] \,.
\end{eqnarray}
This includes the spin-isospin degeneracy factor of the $k$-th particle species $g_k$ as well as its chemical potential $\mu_k$. The Breit-Wigner attenuation in Eq.~(\ref{EqIII}) accounts for a finite width $\Gamma_k$ of hadron resonances with mass $m_k$. The lower limit $M_k^{Th}$ of the mass integration in Eq.~(\ref{EqIII}) stands for the decay threshold mass of the corresponding hadrons in the dominant channel. The factor 
\begin{equation}
\label{EqV}
{N_k (M_k^{Th})} \equiv \int\limits_{M_k^{Th}}^\infty \frac{d m~\, \Gamma_{k}}{(m-m_{k})^{2}+\Gamma^{2}_{k}/4}
\end{equation}
provides a proper normalization of the mass distribution function. A special comment should be made with respect to the chemical potential $\mu_k$ of the $k-$th particle species. Hadronic systems produced in the collisions at such high energies as under LHC conditions are characterized by the almost perfect symmetry between particles and antiparticles. 
This leads to a vanishing of the corresponding chemical potentials \cite{KAB_Ref4,KAB_Ref5}. 
In Eq.~(\ref{EqIII})  they, however, are kept for the convenience of the calculations, while in our analysis we set $\mu_k=0$.

Having Eqs.~(\ref{EqI}) - (\ref{EqII}) solved for a given value of temperature $T$, the particle number density of the $k$-th hadron species can be found with the help of the thermodynamic identity $\rho_k=\frac{\partial p}{\partial\mu_k}$. 
This number density gives a direct access to the thermal yields of particles $N_k^{th} =V \rho_k$, defined through the the 
effective volume of the system at chemical freeze out (CFO) $V$. Since inelastic reactions cease to occur after CFO, the particle decays are no longer compensated by their inverse processes. 
As a result the total yields of particles $N_k^{tot}$ differ from the thermal ones. 
The corresponding modification can be done using the branching ratios $Br_{l\rightarrow k}$, i.e. the probability of a particle $l$ to decay strongly into a particle $k$. The total particle multiplicities allow us to define the particle yield ratios, which are independent on the CFO volume, i.e.
\begin{equation}
\label{EqVI}
\frac{N^{tot}_k}{N^{tot}_j}=
\frac{\rho_k+\sum_{l\neq k}\rho_l\, Br_{l\rightarrow k}}{\rho_j+
\sum_{l\neq j}\rho_l \, Br_{l\rightarrow j}}\,.
\end{equation}
These ratios can be compared to the experimental data, while the temperature should be adjusted in order to provide the best agreement. In our analysis we used the values of hadronic hard-core radii found in previous works, while the hard-core radius of the sexaquark was analyzed independently.
More details on the fitting procedure of experimental data with the HRGM can be found in \cite{KAB_Ref4,KAB_Ref5,KAB_Ref6new}.

\section{Results}
\label{section_results}

In this paper we 
 show  results on the ratio of thermal sexaquark yields to thermal deuteron  ($d$) yields and to thermal Omega ($\Omega$) hyperon yields.
This comparison aims to address two distinct cases, in particular in order to study 
two different  expected times of  freeze
out, for each of the $\Omega$ hyperon and the deuteron, which is related to  their  internal structure.

Namely we consider  the case of the nuclei, like the deuteron which are expected in principle to build
at a late stage of the collision through coalescence of nucleons and secondly, 
the
case of the $\Omega$ hyperon which is expected to have a chemical freeze out at an earlier 
stage of the evolution of the hadronization \cite{KAB_Ref30new}.
 We consider these two cases as kind of two distinct freeze out assumptions in which the sexaquark production can be 
assumed.

\begin{figure}
\begin{center}
\includegraphics[width=12.0cm] {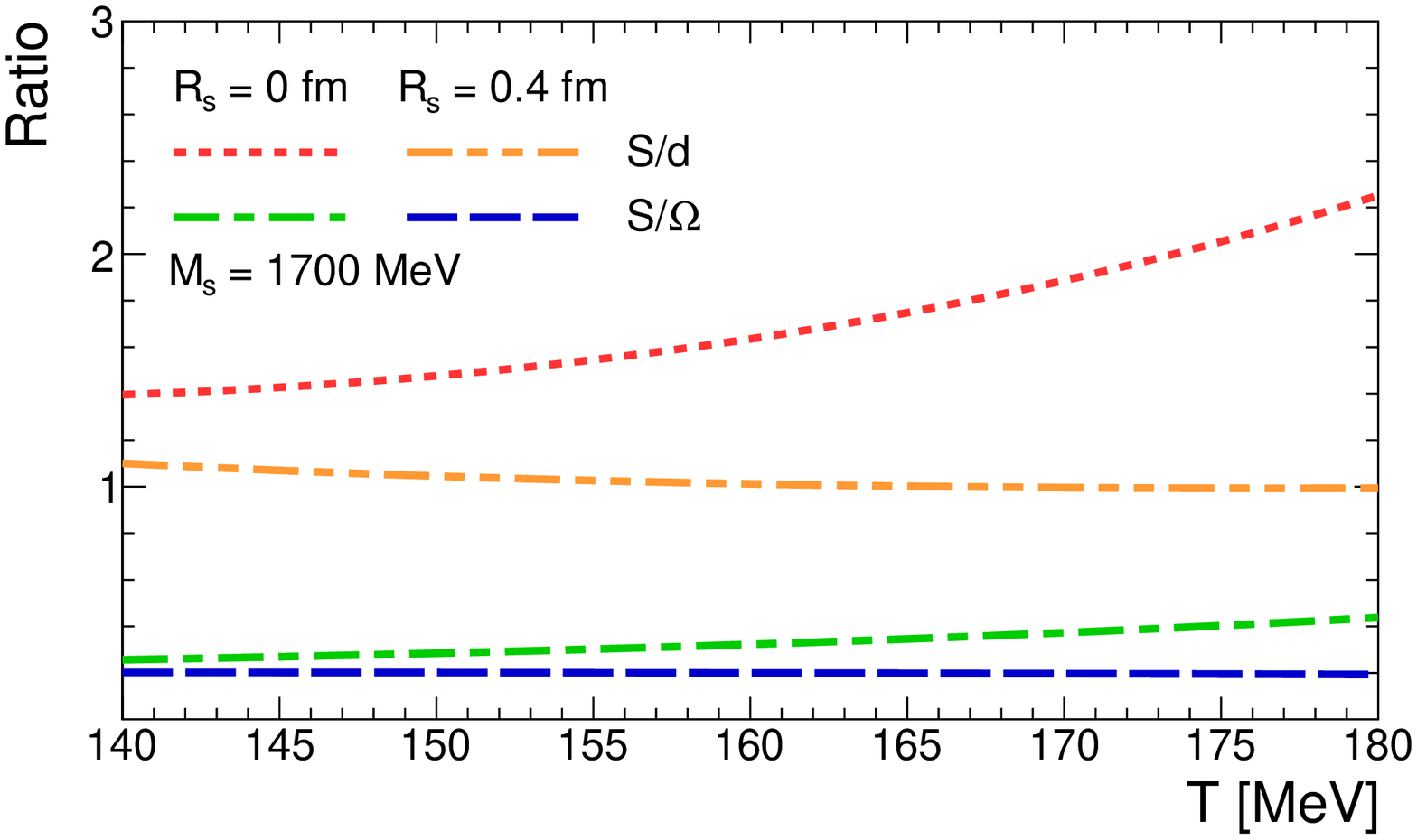}
\\
\includegraphics[width=12.0cm] {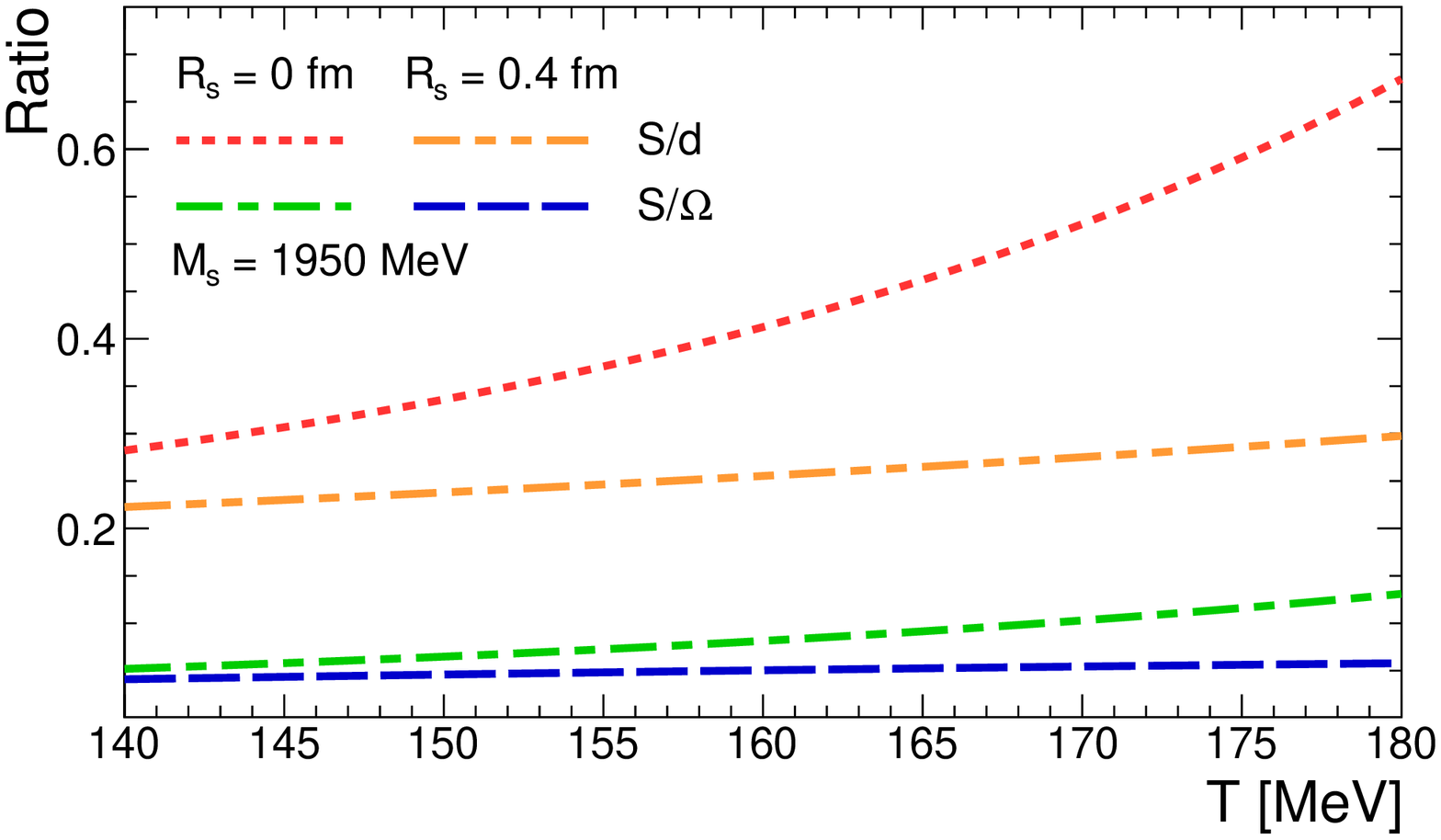}
\caption{Ratio of thermally produced sexaquarks
to thermally produced  hadrons and nuclei, as a function of
the temperature of their thermal particle source, assuming
 the mass of the sexaquarks to be 1700 MeV (upper picture) and 1950 MeV (lower picture).
The two upper lines show the ratio of thermal sexaquarks to thermal deuterons,
whereas the two lower lines show the ratio of thermal sexaquarks to thermal $\Omega$ hyperons.
The $R$ parameter shown on the lines is the assumed hard-core radius of the sexaquark.
}
\label{fig1}
\end{center}
\end{figure}

In figure \ref{fig1} we show the result of the thermal model calculation for the production of sexaquarks in Pb+Pb collisions at the LHC at CERN for a center-of-mass energy of $\sqrt{s_{NN}}$=2.76 TeV.
The $y$-axis shows the ratio of the thermal production yield of sexaquarks to the thermal production yield of selected hadrons and nuclei, as a function of the temperature of the thermal particle source, assuming the mass of the sexaquark to be 1700 MeV and 1950 MeV.
We will explore a larger range of masses in future work.

In both figures, the upper two lines show the ratio of thermal sexaquark yields to thermal deuteron yields, and the two lower lines show the ratio of thermal sexaquark yields to thermal  $\Omega$ hyperon yields.
The variable denoted by $R$ is the assumed hard-core radius of the sexaquark.
From top to bottom, each line corresponds to  $R=0$, $0.4$, $0$, $0.4$ fm. The lines corresponding to $R=0$ fm case show the thermal model result for point-like particles.
We will explore a larger range of $R$ values an sexaquark masses in future work.

The range of temperatures studied here is between 140 and 180 MeV, which is a relevant range for the temperature of the QCD phase transition  that is expected to have occurred in the early universe and can be reproduced in small scale today in the laboratory when colliding high-energy heavy ions in accelerators.
The particle source in Pb+Pb collisions at the center-of-mass energy of $\sqrt{s_{NN}}=2.76$ TeV at LHC
was found in previous papers analysing hadron production in these reactions to be characterized by a temperature of  about 150 MeV.

We find that the ratio of the thermal sexaquarks to thermal 
light nuclei and strange hadrons varies between about 0.2 and 1.4 
for the sexaquark mass of 1700 MeV, and between
about 0.05 and 0.27 MeV for the sexaquark mass of  1950 MeV.
In particular the ratio of the thermal sexaquark yields to thermal deuteron yields is about 1.1 to 1.4
for the sexaquark mass of 1700 MeV, and between
0.22 to 0.27 for
for the sexaquark mass of 1950 MeV.
 The ratio of the thermal sexaquark yields to thermal $\Omega$ hyperon yields is about 0.2 to 0.25 
for the sexaquark mass of 1700 MeV, and about 0.05 
for the sexaquark mass of 1950 MeV.
The ratio to thermal $\Omega$ hyperon yields is rather similar for pointlike and non-pointlike cases
as compared to the ratio to the thermal deuteron yields, especially at lower temperatures.

Furthermore, we observe that the dependence of the ratio of the thermal sexaquark yields to thermal light nuclei and strange hadron yields has a weak dependence on the temperature except for the case of  pointlike particles (R=0) in which case the ratio rises with temperature.
The expected chemical freeze-out temperature of the Pb+Pb system at the center-of-mass energy of $\sqrt{s_{NN}}$=2.76 TeV
is about 150 MeV.
For the sexaquark mass of 1700 MeV and at the temperature of 150 MeV,
 the ratio of the thermal  sexaquark yields to the thermal  light nuclei yields is around 1.45 for $R=0$ fm and around 1.00 for $R=0.4$ fm, while the ratio of the thermal  sexaquark yields to thermal $\Omega$ hyperon yields is around 0.25 for $R=0$ fm and around $0.20$ for $R=0.4$ fm.
For the sexaquark mass of 1950 MeV and at the temperature of 150 MeV,
 the ratio of the thermal  sexaquark yields to the thermal  light nuclei yields is around 0.27 for $R=0$ fm and around 0.22 for $R=0.4$ fm, while the ratio of the thermal  sexaquark yields to thermal $\Omega$ hyperon yields is around 0.05 for both $R=0$ fm and around $0.04$ for $R=0.4$ fm.

In all cases the ratio of the thermal  sexaquark yields to thermal light nuclei and strange hadron yields is quite high and therefore abundant production of thermal sexaquarks  may be expected
in heavy-ion collisions at LHC energies.

\section{Conclusions and Outlook}
\label{section_conclusions}

A stable, compact $uuddss$ state has been proposed as a dark matter candidate by one of us (G. Farrar) and was named sexaquark in order to make clear the difference from previous
theoretical works on a $uuddss$ state with different characteristics that was proposed as $H0$ dibaryon by R. Jaffe. 
If indeed there is a stable sexaquark and the amplitude for its breakup to two baryons is small as argued, then the  experimentally measured ratio of 
Omega(dark matter)/Omega(matter) in the Universe can be very well reproduced, suggesting that $(u,d,s)$-symmetric matter of any form may be the source  of dark matter in the Universe.

In this work we estimate for the first time the thermal production rates of sexaquarks and in particular the ratios of thermal sexaquark states to thermal hadrons with the particular example of the $\Omega$ baryon, as well as to thermal light nuclei with the particular example of the deuteron, in heavy-ion collisions at a center-of-mass energy of $\sqrt{s_{NN}}=2.76$ TeV at the LHC.
The study has been performed an assumed mass of 1700 and 1950 MeV for the sexaquark and for 2 cases of radii of the sexaquarks, $R=0.0$ and 0.4 fm.

It is shown that
sexaquarks may be  produced at relatively high rates for both cases of the investigated radii of 0.0 and 0.4 fm.
At 150 MeV, thermal sexaquarks with a mass of 1700 MeV are found to be produced with similar rate as thermal deuterons,
and  thermal sexaquarks with a mass of 1950 MeV 
are found to be produced with a rate of 1/4 of the thermal deuterons,
 in Pb+Pb collisions at the center-of-mass energy of $\sqrt{s_{NN}}=2.76$ TeV at the LHC.

The estimated production rate of the sexaquark in heavy-ion collisions is of great interest for future searches of this hypothetical particle and may be relevant for other considerations about the role of the sexaquark in cosmological phase transitions and in neutron stars.
We will explore a larger range of masses, hard-core radii $R$,
collision systems and center of mass collision energies in future work.

\section{Acknowledgments}
The work of D.B. and O.I. was supported by the Polish National Science Center under grant No. 2019/33/B/ST9/03059. 
D.B. received funding from the Russian Foundation for Basic Research (RFBR) under grant No. 18-02-40137. 
K.B., O.I. and V.S. acknowledge 
partial support by the Section of Physics and Astronomy of the National 
   Academy of Science  of Ukraine under Grant No. 0118U003197.
V.S. is also thankful for the support by the Funda\c c\~ao para a Ci\^encia e Tecnologia (FCT), Portugal, under the project No. UID/04564/2021.
L.B. and E.Z. acknowledge support by the Norwegian Research Council (NFR) under grant No. 255253/F50, “CERN Heavy Ion Theory”, and by the RFBR under grants No. 18-02-40084
and No. 18-02-40085.
G.R.F. was supported in part by NSF-2013199.
I.K.P. was supported in part by grants ANID - Chile FONDECYT 1170319, and ANID PIA/APOYO AFB180002.
S.Kabana acknowledges partial support by ANID PIA/APOYO AFB180002 (Chile).
S.Kuleshov and S.Kabana acknowledge
partial support by the Millennium Institute for SubAtomic Physics at the High-energy FrontIer (SAPHIR),
\verb|ICN2019_044|, ANID, Chile.
S.Kuleshov acknowledge partial support by the grant FONDECYT 1191103, ANID, Chile.
A.T  received funding by Russian Federation for Basic Research (RFBR) under grant No. 18-02-40086 and was supported by the Ministry of Science and Higher Education of the Russian Federation, Project “Fundamental properties of elementary particles and cosmology” No. 0723-2020-0041.

\end{document}